\documentclass[twocolumn,superscriptaddress,showpacs,amssymb,amsmath,amsfont
s,aps]{revtex4}

\usepackage{graphicx}
\usepackage{dcolumn}
\usepackage{bm}

\begin{document}


\title{Search for the $\Theta^+$ pentaquark in the $\gamma d \to \Lambda n K^+$ reaction measured with CLAS} 
\newcommand*{\ORSAY}{Institut de Physique Nucl\'eaire, Orsay, France}
\affiliation{\ORSAY}
\newcommand*{\INFNFR}{INFN, Laboratori Nazionali di Frascati, 00044 Frascati, Italy}
\affiliation{\INFNFR}
\newcommand*{\SCAROLINA}{University of South Carolina, Columbia, South Carolina 29208}
\affiliation{\SCAROLINA}
\newcommand*{\OHIOU}{Ohio University, Athens, Ohio  45701}
\affiliation{\OHIOU}
\newcommand*{\ECOSSEG}{University of Glasgow, Glasgow G12 8QQ, United Kingdom}
\affiliation{\ECOSSEG}
\newcommand*{\JLAB}{Thomas Jefferson National Accelerator Facility, Newport News, Virginia 23606}
\affiliation{\JLAB}
\newcommand*{\ANL}{Argonne National Laboratory, Argonne, IL 60439}
\affiliation{\ANL}
\newcommand*{\ASU}{Arizona State University, Tempe, Arizona 85287-1504}
\affiliation{\ASU}
\newcommand*{\UCLA}{University of California at Los Angeles, Los Angeles, California  90095-1547}
\affiliation{\UCLA}
\newcommand*{\CSU}{California State University, Dominguez Hills, Carson, CA 90747}
\affiliation{\CSU}
\newcommand*{\CMU}{Carnegie Mellon University, Pittsburgh, Pennsylvania 15213}
\affiliation{\CMU}
\newcommand*{\CUA}{Catholic University of America, Washington, D.C. 20064}
\affiliation{\CUA}
\newcommand*{\SACLAY}{CEA-Saclay, Service de Physique Nucl\'eaire, F91191 Gif-sur-Yvette, France}
\affiliation{\SACLAY}
\newcommand*{\CNU}{Christopher Newport University, Newport News, Virginia 23606}
\affiliation{\CNU}
\newcommand*{\UCONN}{University of Connecticut, Storrs, Connecticut 06269}
\affiliation{\UCONN}
\newcommand*{\ECOSSEE}{Edinburgh University, Edinburgh EH9 3JZ, United Kingdom}
\affiliation{\ECOSSEE}
\newcommand*{\FIU}{Florida International University, Miami, Florida 33199}
\affiliation{\FIU}
\newcommand*{\FSU}{Florida State University, Tallahassee, Florida 32306}
\affiliation{\FSU}
\newcommand*{\GWU}{The George Washington University, Washington, DC 20052}
\affiliation{\GWU}
\newcommand*{\ISU}{Idaho State University, Pocatello, Idaho 83209}
\affiliation{\ISU}
\newcommand*{\INFNGE}{INFN, Sezione di Genova, 16146 Genova, Italy}
\affiliation{\INFNGE}
\newcommand*{\ITEP}{Institute of Theoretical and Experimental Physics, Moscow, 117259, Russia}
\affiliation{\ITEP}
\newcommand*{\JMU}{James Madison University, Harrisonburg, Virginia 22807}
\affiliation{\JMU}
\newcommand*{\KYUNGPOOK}{Kyungpook National University, Daegu 702-701, Republic of Korea}
\affiliation{\KYUNGPOOK}
\newcommand*{\UMASS}{University of Massachusetts, Amherst, Massachusetts  01003}
\affiliation{\UMASS}
\newcommand*{\MOSCOW}{Moscow State University, General Nuclear Physics Institute, 119899 Moscow, Russia}
\affiliation{\MOSCOW}
\newcommand*{\UNH}{University of New Hampshire, Durham, New Hampshire 03824-3568}
\affiliation{\UNH}
\newcommand*{\NSU}{Norfolk State University, Norfolk, Virginia 23504}
\affiliation{\NSU}
\newcommand*{\ODU}{Old Dominion University, Norfolk, Virginia 23529}
\affiliation{\ODU}
\newcommand*{\RPI}{Rensselaer Polytechnic Institute, Troy, New York 12180-3590}
\affiliation{\RPI}
\newcommand*{\RICE}{Rice University, Houston, Texas 77005-1892}
\affiliation{\RICE}
\newcommand*{\URICH}{University of Richmond, Richmond, Virginia 23173}
\affiliation{\URICH}
\newcommand*{\UNIONC}{Union College, Schenectady, NY 12308}
\affiliation{\UNIONC}
\newcommand*{\VT}{Virginia Polytechnic Institute and State University, Blacksburg, Virginia   24061-0435}
\affiliation{\VT}
\newcommand*{\VIRGINIA}{University of Virginia, Charlottesville, Virginia 22901}
\affiliation{\VIRGINIA}
\newcommand*{\WM}{College of William and Mary, Williamsburg, Virginia 23187-8795}
\affiliation{\WM}
\newcommand*{\YEREVAN}{Yerevan Physics Institute, 375036 Yerevan, Armenia}
\affiliation{\YEREVAN}
\newcommand*{\NOWUNH}{University of New Hampshire, Durham, New Hampshire 03824-3568}
\newcommand*{\NOWUMASS}{University of Massachusetts, Amherst, Massachusetts  01003}
\newcommand*{\NOWMIT}{Massachusetts Institute of Technology, Cambridge, Massachusetts  02139-4307}
\newcommand*{\NOWCUA}{Catholic University of America, Washington, D.C. 20064}
\newcommand*{\NOWECOSSEE}{Edinburgh University, Edinburgh EH9 3JZ, United Kingdom}
\newcommand*{\NOWGEISSEN}{Physikalisches Institut der Universitaet Giessen, 35392 Giessen, Germany}
\newcommand*{\SAOPAULO}{Physics Institute, University of Sao Paulo, Sao Paulo, Brazil}
\author {S.~Niccolai}
\affiliation{\ORSAY}
\author{M.~Mirazita}
\affiliation{\INFNFR}
\author{P.~Rossi}
\affiliation{\INFNFR}
\author {N.A.~Baltzell} 
\affiliation{\SCAROLINA}
\author {D.S.~Carman} 
\affiliation{\OHIOU}
\author {K.~Hicks} 
\affiliation{\OHIOU}
\author {B.~McKinnon}
\affiliation{\ECOSSEG}
\author {T.~Mibe} 
\affiliation{\OHIOU}
\author {S.~Stepanyan} 
\affiliation{\JLAB}
\author {D.J.~Tedeschi} 
\affiliation{\SCAROLINA}
% ...
\collaboration{The CLAS Collaboration}
     \noaffiliation

\date{\today}

\begin{abstract}
\noindent For the first time, the reaction $\gamma d \to \Lambda n K^+$ has been analyzed in order to search for the exotic pentaquark baryon $\Theta^+(1540)$. The data were taken at Jefferson Lab, using the Hall-B tagged-photon beam of energy between 0.8 and 3.6 GeV and the CEBAF Large Acceptance Spectrometer (CLAS). No statistically significant structures were observed in the $nK^+$ invariant mass distribution. The upper limit on the $\gamma d \to \Lambda \Theta^+$ integrated cross section has been calculated and found to be between 5 and 25 nb, depending on the production model assumed. The upper limit on the differential cross section is also reported.
\end{abstract}

\maketitle

Since the first publication of the observation of the new state $\Theta^+(1540)$ in the year 2003 \cite{nakano}, the possible existence of exotic baryons that have quantum numbers which require a minimum quark content of $qqqq\bar{q}$ has generated tremendous interest in the physics community.
Although the idea of exotic pentaquark states was introduced originally in the early 70's, the specific prediction for both a mass of 1530 MeV/{\it c}$^2$  and a narrow width of less than 15 MeV/{\it c}$^2$, which motivated the first measurement at LEPS/SPring-8 \cite{nakano}, was made in 1997 by Diakonov {\it et al.} \cite{diakonov}. Within the framework of their Chiral Soliton Model, they predicted the $\Theta^+$ to be an isosinglet member of a $J=\frac{1}{2}^+$ anti-decuplet of pentaquark states, having an exotic flavor quantum number $S(\Theta^+)=+1$ and a minimal quark content of $uudd\bar{s}$.

Experimental evidence for the $\Theta^+$ state has been claimed in several published works \cite{nakano,barmin,stepan,kuba,barth,airapetian,asratyan,aleev,abdel,chekanov}. The observation of a candidate for the anti-charmed equivalent of the $\Theta^+$ ($uudd\bar{c}$), of mass 3.1 GeV/{\it c}$^2$, has been claimed by the H1 Collaboration \cite{h1}.
One experiment \cite{alt} also reported the observation of two other pentaquarks, $\Xi^{--}_5$ and $\Xi^{0}_5$. However, several reports of non-observation of pentaquarks have also been made \cite{atkas,schael,aubert,abe,bai,gorelov,stenson,abt,longo,napolitano,armstrong,pinkerburg,antipov,adamovich,g11,mckinnon,airapetian2,compass}. Moreover, the statistical significances of the observed $\Theta^+$ signals are rather low, and there are discrepancies in the measured masses. 
The still-open question of the existence of narrow five-quark baryons can therefore be addressed only by performing a second generation of dedicated, high-statistics experiments. The CLAS Collaboration is currently pursuing high-statistics searches for the $\Theta^+$ through photoproduction on hydrogen \cite{battaglieri} and deuterium \cite{hicks} targets, and in various final states.

\begin{figure} 
\includegraphics[scale=0.5]{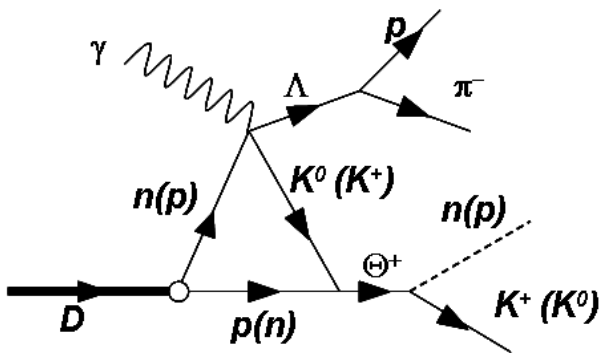}
\caption{\label{reaction} A possible reaction mechanism for the photoproduction of $\Lambda \Theta^+$ on the deuteron.}
\end{figure}

Searching for the $\Theta^+$ through photoproduction on the deuteron together with a $\Lambda$ hyperon has various experimental advantages. The main advantage of this reaction channel is that there are no competing channels to remove in the final state, while at the same time it excludes kinematical reflections of heavy mesons in the $NK$ invariant mass spectrum, possible in other channels, like $pK^+K^-n$ or $pK^+K^0p$ \cite{djerba}. Moreover, the presence of the $\Lambda$ provides a ``strangeness tag'' ($S_{\Lambda}=-1$) in both the $nK^+$ and the $pK^0$ decay modes. 
Figure ~\ref{reaction} shows a possible diagram that could lead to $\Theta^+$ production via a two-step process. The photon interacts with one of the nucleons in the deuteron and produces a $\Lambda$ and a kaon. The $\Lambda$ leaves the target nucleus, while the $K$ rescatters on the spectator nucleon to form a $\Theta^+$. The rescattering probability is determined by the deuteron wave function and the $KN$ scattering cross section. This kind of process has been taken into account by Guzey \cite{guzay} to calculate the total and differential cross section for the $\gamma d \to \Lambda \Theta^+$ reaction. Also, calculations of the $KN$ rescattering amplitude and the probability of production of a narrow resonant state have been performed by Laget \cite{laget}.

We have searched for the $\Theta^+$ in the $\gamma d \to \Lambda nK^+$ reaction using the CLAS-G10 data set \cite{hicks,mckinnon} (the results of our analysis on the $pK^0$ mode, currently underway, will be presented in an upcoming publication). The data were taken during spring 2004 with the Hall-B tagged-photon beam \cite{sober} of energy between 0.8 GeV and 3.6 GeV, impinging on a 24-cm-long liquid-deuterium target. Two different values for the CLAS torus magnetic field \cite{mecking} were chosen for the two halves of the experiment. The run with lower magnetic field had higher acceptance for negative particles in the forward direction, and has been used for this analysis. For this part of the experiment, an integrated luminosity of 31 pb$^{-1}$ has been achieved. The run with higher magnetic field was taken to reproduce the same acceptance and track resolution of the data used for the CLAS published result on the deuteron in the $pK^+K^-n$ channel \cite{stepan}, but had very low acceptance for the $\gamma d \to \Lambda nK^+$ reaction, giving approximately a factor of six less statistics than the low-field run, and was therefore not used for the analysis discussed in this Letter.

\begin{figure} 
\includegraphics[scale=0.4]{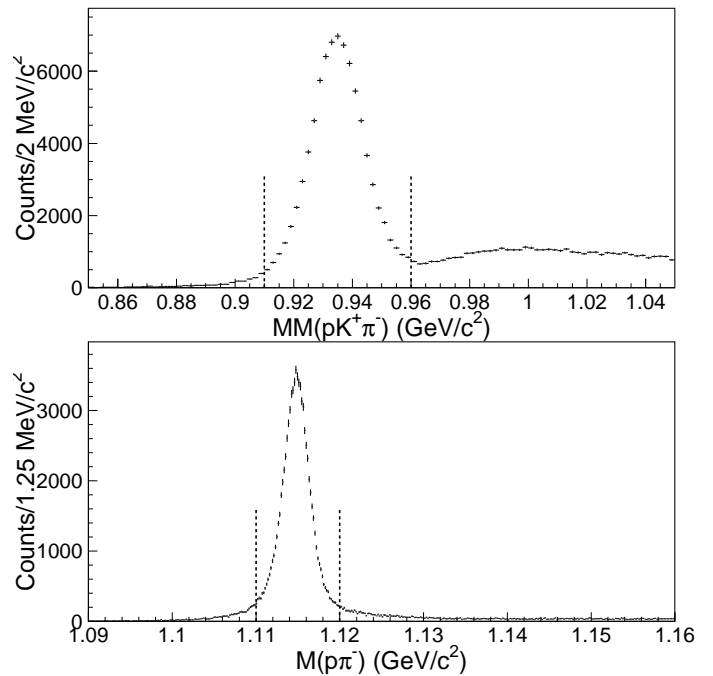}
\caption{\label{fig_mass} Top plot: missing mass of $\gamma d \to p\pi^-K^+X$, showing a peak at the neutron mass. The particle identification cuts for the three charged particles and the selection cut on the $\Lambda$ have been applied. Bottom plot: invariant mass of the $p\pi^-$ system, showing a peak at the $\Lambda$ mass. The particle identification cuts for the three charged particles and the selection cut on the neutron mass have been applied. For both plots, the vertical dotted lines represent the $3\sigma$ selection cuts (with $\sigma_n = 0.009$ GeV/{\it c}$^2$ and $\sigma_{\Lambda} = 0.002$ GeV/{\it c}$^2$) applied to select the final state.}
\end{figure}

Since CLAS is mainly efficient for the detection of charged particles, the $\Lambda \to p \pi^-$ decay mode was chosen. The final state was determined exclusively, identifying the 3 charged particles ($p, \pi^-, K^+$) through their momenta and times of flight measured in CLAS, reconstructing the neutron with the missing mass technique (Fig.~\ref{fig_mass}, top plot), and the $\Lambda$ via the $p \pi^-$ invariant mass (Fig.~\ref{fig_mass}, bottom plot). Selection cuts $3\sigma$-wide were placed around both the neutron peak in the missing mass and the $\Lambda$ peak in the invariant mass, as shown by the dotted lines in Fig.~\ref{fig_mass}.

\begin{figure} 
\includegraphics[scale=0.35]{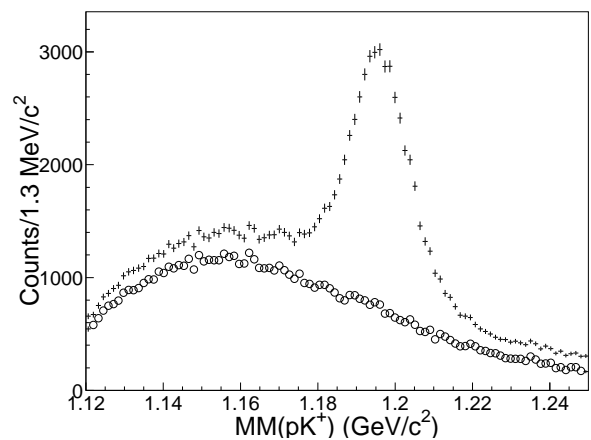}
\caption{\label{sigma_lf} Missing mass of $pK^+$ before (crosses) and after (circles) applying the $\Lambda$ selection cut. The $\Sigma^-$ signal, visible before applying the $\Lambda$ cut, is eliminated when the $\Lambda nK^+$ events are selected.}
\end{figure}

The $p \pi^- n K^+$ final state can also arise from the $\gamma d \to \Sigma^- p K^+$ channel, when the $\Sigma^-$ decays weakly into $n\pi^-$. In order to study this possible source of background, the distribution of the missing mass of the $pK^+$ system has been studied. As expected, the $\Sigma^-$ peak (Fig.~\ref{sigma_lf}, crosses) disappears after applying the $\Lambda$ selection cut on the $p\pi^-$ invariant mass (circles). 

\begin{figure} 
\includegraphics[scale=0.48]{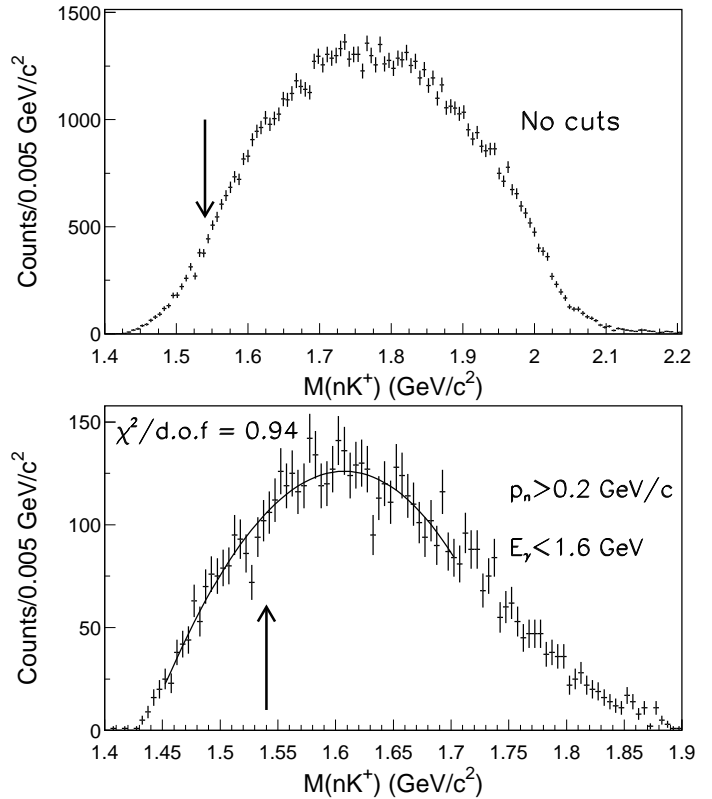}
\caption{\label{fig_nk} Raw distributions of the invariant mass of the $nK^+$ system after channel selection. Top plot: no kinematical cuts are applied. Bottom plot: the $E_{\gamma}<1.6$ GeV and $p_n>0.2$ GeV/{\it c} kinematical cuts are applied. No statistically significant structure is visible in the mass range around 1.54 GeV/{\it c}$^2$, indicated by the arrows. The third-order polynomial fit used for the upper limit estimate is shown.}
\end{figure}

After selecting the $\Lambda nK^+$ events, the $\Theta^+$ signal was searched for in the invariant mass of the $nK^+$ system. The result obtained is shown in the top plot of Fig.~\ref{fig_nk}. Since the $nK^+$ mass spectrum does not show any evident structure, the following kinematical cuts were subsequently imposed based om the model of Ref.~\cite{guzay} in order to try to enhance a possible $\Theta^+$ signal over the non-resonant $nK^+$ background:
\begin{itemize}
\item{non-spectator-neutron cuts: the non-resonant $nK^+$ background can be suppressed by removing the events in which the neutron is a spectator, having momentum given by the Fermi-momentum distribution in the deuteron.}
\item{photon-energy cuts: according to the model \cite{guzay}, the $\gamma d \to \Lambda \Theta^+$ cross section decreases rapidly with increasing photon energy.}
\end{itemize}
Several cuts on the neutron momentum ($p_n$) and on the photon energy ($E_{\gamma}$) have been tried. However, also under these stringent kinematic conditions, no narrow peaks having statistical relevance can be observed in the mass region around 1.54 GeV/{\it c}$^2$. An example is given in the bottom plot of Fig.~\ref{fig_nk}, where the kinematic requirements $p_n>0.2$ GeV/{\it c} and $E_{\gamma}<1.6$ GeV are applied. A parallel analysis based upon a kinematic fitting procedure was also carried out, leading to equivalent final results.

\begin{figure} 
\includegraphics[scale=0.48]{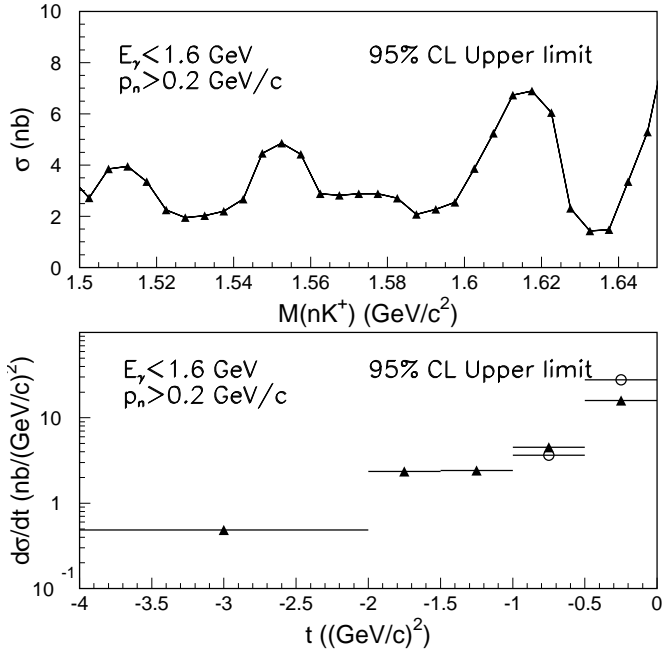}
\caption{\label{fig_ul} Top: upper limit (at the 95\% confidence level) on the $\gamma d \to \Lambda \Theta^+$ total cross section as a function of the $nK^+$ invariant mass. Model (a) has been used to extract the integrated acceptance. Bottom: upper limit of the differential cross section $d\sigma/dt$ as a function of $t$, for $1.52<M(nK^+)<1.56$ GeV/{\it c}$^2$. The triangles and the circles represent, respectively, the results obtained using model (a) and (c).}
\end{figure}

Since no structures having relevant statistical significance appear in the $nK^+$ invariant mass for any of the kinematic cuts that have been studied, the upper limit on the cross section has been calculated for $p_n>0.2$ GeV/{\it c} and $E_{\gamma}<1.6$ GeV.
For each bin in $M(nK^+)$, the number of events above the background was calculated as follows: the $nK^+$ distribution was fitted with a third-order polynomial (as shown in the bottom plot of Fig.~\ref{fig_nk}), and then a second fit was performed by fixing the third-order polynomial and adding a Gaussian curve having a fixed centroid at the $M(nK^+)$ bin under examination and a width equal to 5 MeV/{\it c}$^2$. This width corresponds to the invariant-mass resolution of CLAS determined via Monte-Carlo simulations. Only the amplitude of the Gaussian was left as a free parameter for the fit. The yield above or below the curve describing the background is therefore given by the integral of the Gaussian. The upper limit at the 95\% confidence level on the yield was calculated using the Feldman-Cousins method \cite{feldman}. The acceptance has been computed with the aid of a Monte-Carlo simulation reproducing the response of CLAS, with three different models used to generate the $\Lambda n K^+$ final state: (a) a two-body ($\Lambda \Theta^+$) phase space, followed by the decay $\Theta^+ \to n K^+$, with an energy-independent cross section and a bremsstrahlung photon-energy distribution; (b) a $\Lambda n K^+$ final state for which the kinematical variables are tuned to match the experimental data; and (c) a two-body ($\Lambda \Theta^+$) final state based upon the model of Guzey \cite{guzay}, followed by the decay $\Theta^+ \to n K^+$. The integrated acceptances obtained with model (a) and model (b) are comparable and are of the order of 0.5\%.  Model (c) produces most of the $\Lambda$'s (i.e. $\pi^-$'s) in the very forward direction, where CLAS has no acceptance for negative particles, and thus it gives an integrated acceptance about a factor of 5 smaller than for models (a) and (b). Therefore, the integrated acceptance is strongly model dependent. The $\Lambda \to p\pi^-$ decay branching ratio (64\%) was included in the calculation of the acceptance, as well as the $\Theta^+$ decay branching ratio for the $nK^+$ mode, which was assumed to be 50\%. The photon flux was measured by integrating the tagged-photon rate during the data-acquisition livetime. The tagging efficiency was measured during dedicated low-flux runs, using a lead-glass total-absorption detector \cite{sober}.
 The resulting upper limit on the $\gamma d \to \Lambda \Theta^+$ total cross section is shown, as a function of $M(nK^+)$, in the top plot of Fig.~\ref{fig_ul}. In the mass range between 1.52 and 1.56 GeV/{\it c}$^2$ the upper limit is 5 nb. Here, the acceptance obtained with model (a) has been used. Adopting model (c) to extract the total cross section gives an upper limit about a factor of 5 larger than the one shown in Fig.~\ref{fig_ul}.

The upper limit on the $\gamma d \to \Lambda \Theta^+$ differential cross section as a function of the momentum transfer $t$, with $t=(p^{\mu}_{\gamma}-p^{\mu}_{\Lambda})^2$, has also been calculated, again for $p_n>0.2$ GeV/{\it c} and $E_{\gamma}<1.6$ GeV.
The data were divided into five $t$ bins, as shown in the lower plot of Fig.~\ref{fig_ul}. For each $t$ bin, the upper limit on the cross section was extracted according to the procedure described above for the total cross section, using the acceptances given by models (a) (triangles) and (c) (circles).
The maximum value of the upper limit in the $M(nK^+)=1.52-1.56$ GeV/{\it c}$^2$ range for each $t$ bin was then used to get the upper limit on the differential cross section, as shown in the bottom plot of Fig.~\ref{fig_ul}. It varies between 0.5 nb/(GeV/{\it c})$^2$ at the highest values of $-t$ and 30 nb/(GeV/{\it c})$^2$ as $t$ approaches 0. The kinematic region at small $t$ values, however, corresponds to the forwardmost part of the spectrometer, where the acceptance drops to zero. This explains the higher value on the upper limit for the last bin in Fig.~\ref{fig_ul}. 

In conclusion, for the first time a search for the exotic pentaquark $\Theta^+$ in the $\gamma d \to \Lambda n K^+$ reaction was performed. The high-statistics CLAS-G10 data were used for this search, and the final state was cleanly identified. No statistically significant signal was observed in the $nK^+$ invariant mass distribution, even under several different kinematic conditions. Upper limits on the total cross section were calculated in the mass range between 1.52 and 1.56 GeV/{\it c}$^2$ and for $p_n>0.2$ GeV/{\it c} and $E_{\gamma}<1.6$ GeV, and found to be 5 nb when computed with the phase-space Monte-Carlo acceptance, while this number increases by a factor of 5 if the Guzey model is used. The upper limit on the differential cross section as a function of $t$ has also been extracted and found to be between 0.5 and 30 nb/(GeV/{\it c})$^2$.

We would like to thank the staff of the Accelerator and Physics
Divisions at Jefferson Lab, who made this experiment possible. Acknowledgments
for the support of this experiment go also to the Italian Istituto
Nazionale di Fisica Nucleare, the French Centre National de la
Recherche Scientifique and Commissari\'at a l'Energie Atomique, the UK Engineering and Physical Science Research Council, the
U.S. Department of Energy and the National Science Foundation, and the
Korea Research Foundation. The Southeastern Universities
Research Association (SURA) operates the Thomas Jefferson National
Accelerator Facility under U.S. Department of Energy contract
DE-AC05-84ER40150.


\end{document}